# Angular dependence of the electrically driven and detected ferromagnetic resonance in Ni$_{36}$Fe$_{64}$ wires

Qiang Gao and Maxim Tsoi

*Department of Physics, University of Texas at Austin, Austin, TX 78712, USA*

*Texas Materials Institute, University of Texas at Austin, Austin, TX 78712, USA*

We study the angular dependence of ferromagnetic resonance (FMR) in Ni$_{36}$Fe$_{64}$ wires using both traditional microwave-absorption and electrical-detection techniques. In our experiments we apply a static magnetic field at an angle $\theta$ with respect to the wire, while the microwave current, which is responsible for driving FMR, is always flowing along the wire. For different $\theta$s we find a very similar behavior for both microwave-absorption and electrically-detected FMR – the resonance magnetic field follows a simple "1/cos($\theta$)" dependence. This simple behavior highlights the importance of the relative orientation between the driving current and magnetic field. We also investigated the dependence of the electrically detected FMR on dc and rf (microwave) current magnitudes. As expected, the resonance signal increases linearly with both the applied dc current and the microwave power.

## I. Introduction

Ferromagnetic resonance (FMR) is a powerful method to study magnetic dynamics in various media, from bulk magnetic materials to nanoscale heterostructures [1]. The resonance occurs when the natural precession frequency of magnetization in the media matches the frequency of an externally applied rf magnetic field. The latter is often generated by an applied rf (microwave) current flowing through the media. That makes the direction of this driving current an important parameter in FMR experiments.

In this paper we study the dependence of FMR on the relative orientation between the driving current and applied magnetic field by two detection techniques. First, FMR is detected traditionally by measuring the absorption of applied microwaves. Second, FMR is detected electrically. Here the precession of magnetization driven by microwaves produces variations in the media's resistance via mechanisms like anisotropic magnetoresistance (AMR), anomalous Hall effect (AHE), tunneling magnetoresistance (TMR), or inverse spin-Hall effect (ISHE) [2, 3]. These resistance variations, in turn, produce a rectified dc voltage (photovoltage) that can be detected electrically. The latter enables the electrical detection of FMR, which has been widely used to study magnetization and spin dynamics in magnetic nanostructures over the past decade [4, 5].

## II. Methods

In our experiments we use a Ni$_{36}$Fe$_{64}$ wire with diameter 50 μm (Goodfellow FE025100). The 1.3 mm long wire terminates a coaxial cable used to deliver dc and rf (microwave) currents to the wire. The microwave current produces a circumferential (rf) Oersted field that generates a torque on the sample's magnetization and drives FMR [6, 7]. In our experimental setup both dc (current source and voltmeter) and rf (microwave generator and power sensor) electronics were connected to the wire using a bias tee as schematically shown in Fig. 1. The power sensor (Keysight U2002A) is used to detect the microwave power reflected from the wire. The dc voltmeter (Keithley 2182A) is used to detect the dc voltage across the wire, which includes a small rectified voltage (photovoltage) produced by microwaves. Both the microwave power and dc voltage were measured as a function of external magnetic field (up to 0.7 T) applied at an angle $\theta_H$ with respect to the wire (current direction). The angle was varied from 0-360 degrees.

## III. Modeling

Kittel's model [1] is routinely used to describe FMR in ferromagnetic media of different geometry. Our wire diameter (50 µm) is much larger than the electromagnetic skin depth (~1 µm) [8]. Therefore, it is safe to assume that the microwave current is confined to a thin layer under the wire surface. For an external field applied along the wire, this is equivalent to the case of a thin film in a parallel magnetic field for which the Kittel's resonance condition is [1, 8]:

$$\frac{\omega}{\gamma} = \sqrt{H_{res}(H_{res} + 4\pi M_s)} \qquad (1)$$

where $\omega$ is the rf frequency, $\gamma$ the gyromagnetic ratio, $M_s$ the saturation magnetization, and $H_{res}$ the resonance field. For an external field applied perpendicular to the wire, the resonance condition becomes:

$$\frac{\omega}{\gamma} = H_{res} - 4\pi M_s \qquad (2)$$

Finally, for a static field applied at an arbitrary angle $\theta_H$ to the wire, an elaborated Kittel's model [9] predicts the following dispersion relation between $\omega$ and $H_{res}$:

$$\left(\frac{\omega}{4\pi\gamma M_s}\right)^2 = \left[\left(\frac{H_{res}}{4\pi M_s}\right)\cos(\theta_H - \theta_M) + \cos(2\theta_M)\right]\left[\left(\frac{H_{res}}{4\pi M_s}\right)\cos(\theta_H - \theta_M) - \sin^2(\theta_M)\right] \qquad (3)$$

where $\theta_M$ is the angle between $M_s$ and the wire (see Fig. 1), which can be found from:

$$\left(\frac{H}{4\pi M_s}\right)\sin(\theta_H - \theta_M) = \cos(\theta_H)\sin(\theta_M) \qquad (4)$$

Usually, the magnetization angle $\theta_M$ lags behind the applied magnetic field angle $\theta_H$, except in the two special cases: $\theta_M = \theta_H = 0°$ (Eq. 1) and $\theta_M = \theta_H = 90°$ (Eq. 2) where they are aligned with each other.

Alternatively, we will use a simple '*cos*' model to describe FMR, which highlights the importance of the driving current direction. In our experiment (Fig. 1), the microwave current $I_{rf}$ is flowing along the wire and produces a circumferential (rf) magnetic field $h_{rf}$, which drives FMR and is always perpendicular to the wire. We assume that the parallel pumping is negligible and only the perpendicular (to $h_{rf}$) component of the wire magnetization can be driven into FMR. We further assume that the magnetization is always aligned with the applied magnetic field, i.e., $\theta_M = \theta_H$. Then only the perpendicular to $h_{rf}$ (parallel to wire) component of the magnetic field ($H\cos\theta_H$) will contribute to FMR and the resonance condition reduces to Eq. 1 with $H_{res}$ replaced by $H\cos\theta_H$. Therefore, we refer to such a model as the '*cos*' model.

Figure 2 shows the FMR dispersion relations at different $\theta_H$ predicted by the two models – Kittel's model and '*cos*' model. The dispersions predicted by two models are quite different for magnetic fields larger than the saturation magnetization ($\frac{H_{res}}{4\pi M_s} > 1$). However, at relatively small fields ($\frac{H_{res}}{4\pi M_s} < 1$), the predictions are very close to each other.

The electrical detection of FMR [6, 7, 10] is based on rectification and frequency mixing properties of our wire. Its current-dependent mixing characteristics can be described by assuming that the current $I = I_{dc} + i_{rf}cos(\omega t)$ through the wire has a dc and rf (microwave) components. Then the rectification properties of the wire can be found by expanding the resulting voltage *V(I)* across the wire about the bias current $I_{dc}$. Mixing of a time-dependent component of the wire resistance *R=V/I* with the microwave current $i_{rf}cos(\omega t)$ contributes a dc (photovoltage) term to *V(I)*:

$$V_\omega \sim i_{rf}^2 (d^2V/dI^2)_{I_{dc}} \qquad (5)$$

which in the simplest case is $\sim i_{rf}^2 I_{dc}$ [10] and suggests that the resulting photovoltage should increase linearly with the applied rf power ($\sim i_{rf}^2$) and the applied dc bias current $I_{dc}$.

## IV. Results and Discussion

Figure 3a shows FMR absorption and photovoltage spectra at $\theta_H=0°$ for a constant applied frequency (10.25 GHz) and power (17 dBm) at the source. The green trace is the raw data of the microwave power reflected from the wire as a function of magnetic field. It displays a dip in the power at ±0.1 T which corresponds to the maximum absorption of microwaves in the wire at FMR. The blue trace is the dc photovoltage $V_\omega$ induced by the microwaves ($I_{dc}$= -10 mA). It is the difference between dc voltages across the wire with $V(I_{dc} + i_{rf})$ and without $V(I_{dc})$ microwaves. The voltage signal $V(I_{dc})$ without microwaves is shown in black and combines an AMR peak at zero field and linear magnetoresistance at higher fields. The absorption and photovoltage spectra in Fig. 3a are very similar with some differences at low fields and a higher level of noise in the voltage data. The dip in reflected power at ±0.1 T correlates well with the minimum in photovoltage and can be described by the Kittel's FMR condition at $\theta_H=0°$ (Eq. 1) assuming $4\pi M_s$=1.33 T and $\gamma$=27.8 GHz/T.

Figure 3b shows the FMR dispersion relation between the applied frequency and the resonance magnetic field. Solid symbols show the experimental data for different $\theta_H$ = 0, 30, 40, 50, 60, 74, 80, 84, 86° (color coded). Dashed curves are the '*cos*' model fits. The fitting angles ($\theta$ = 0, 28, 37, 46, 55, 69, 76, 80, 83°) are within a few degrees of $\theta_H$ that is consistent with the experimental accuracy of determining $\theta_H=0°$ (± a few degrees) and may also suggest that the magnetization direction ($\theta_M$) lags behind the applied magnetic field by 2-4°.

We have investigated the dependence of the electrically detected FMR on dc and rf (microwave) current magnitudes. Figure 4 shows the peak FMR photovoltage vs dc current at a fixed rf power *P*=21 dBm (Fig. 4a) and vs rf power at a fixed $I_{dc}$=10 mA (Fig. 4b). The linear fits (dashed lines) confirm that the resonance signal increases linearly with both the applied dc current and the microwave power as expected from Eq. 5.

Figure 5 shows the angular dependence of FMR. 2D gray-scale plots show the FMR absorption (Fig. 5a) and photovoltage (Fig. 5b) spectra as a function of $\theta_H$. Lighter color indicates higher power/voltage. The blue and green lines indicate positions of the $\theta_H=0°$ spectra from Fig. 3a. The red curve is the '*cos*' model fit. Both absorption and photovoltage spectra show very similar behaviors as a function of $\theta_H$. The similarities of the absorption and photovoltage spectra in Fig. 3a and their respective angular dependencies in Figs. 5a and 5b suggest that the electrical detection of FMR is essentially equivalent to the traditional absorption measurements.

Figure 6 highlights the angular dependence of the resonance field. The field increases with increasing $\theta_H$. Both the Kittel's model fit (solid curve in Fig. 6) and the '*cos*' model fit (dashed curve) are consistent with this behavior and display some deviations from the experimental data (red symbols) only for very large angles (close to $\theta_H=\pm90°$). This result suggests that the simple '*cos*' model can capture the essence of the angular dependence by assuming that only the parallel (to the wire) component of magnetic field ($Hcos\theta_H$) contributes to FMR. In contrast, the Kittel's model explains the increase of the resonance field by an opposing demagnetizing field, which appears when the magnetization has a component perpendicular to the wire. In order to test this hypothesis experimentally we have performed FMR measurements in an alternative geometry, where the demagnetizing field is expected to have a minimal effect on the resonance field.

Figure 7 shows the angular dependence of the current-driven FMR in a 0.1 thick Fe foil. The rectangular shape of the foil (1.3 mm × 0.7 mm) was chosen to (i) minimize the demagnetizing effects and (ii) limit the current's spreading. The applied magnetic field was rotated in the plane of the foil (see insert to Fig. 7). The 2D gray-scale plot in Fig. 7 shows the absorption spectra as a function of $\theta_H$ and the '*cos*' model fit (red curve). An increase of the resonance field for angles close to $\theta_H=\pm90°$ is obvious despite a significant broadening of the resonance due to the current spreading (compare with Fig. 5a). In this rectangular geometry the demagnetizing effects are expected to be much smaller than those in the wire, that plays in favor of the '*cos*' model and highlights the importance of the relative orientation between the driving current and magnetic field.

## V. Summary

We have experimentally investigated the angular dependence of the electrically driven FMR in $Ni_{36}Fe_{64}$ wires. Two FMR detection techniques were used: traditional microwave-absorption and electrical detection. Both techniques showed very similar results both in terms of the FMR line shape and the angular dependence of the resonance field. The resonance field was found to increase significantly when the field direction approaches the perpendicular-to-wire geometry ($\theta_H=\pm90°$). We have exploited two models – the Kittel's model and the '*cos*' model – to fit the experimental data. Both models are consistent with experimental observations and display some deviations from the data only for very large angles (close to $\theta_H=\pm90°$). We have performed a test experiment with a rectangular Fe foil to distinguish between the models. The resulting '*cos*'-like angular dependence supports the '*cos*' model and highlights the importance of the relative orientation between the driving current and magnetic field.

# Figure Captions

**Figure 1**. Experimental setup. DC (current source and voltmeter) and RF (microwave generator and power sensor) electronics are connected to the $Ni_{36}Fe_{64}$ wire using a bias tee.

**Figure 2**. FMR dispersion. The comparison of the predictions by the Kittel's model (solid curves) and the '*cos*' model (dashed curves) for different $\theta_H$ from 0-89°.

**Figure 3**. (a) FMR photovoltage (blue) and absorption (green) spectra at $\theta_H=0°$. The black trace shows the dc voltage across the wire without microwaves. (b) FMR dispersion. Experimental data (solid symbols) and corresponding '*cos*' model fits (dashed curves) for different $\theta_H$ (color coded).

**Figure 4**. The peak FMR photovoltage vs (a) dc current at a fixed rf power $P=21$ dBm and (b) vs rf power at a fixed $I_{dc}=10$ mA. The insert shows the same data vs power in dBm. The dashed lines are linear fits.

**Figure 5**. 2D gray-scale plots show the FMR absorption (a) and photovoltage (b) spectra as a function of the magnetic field angle $\theta_H$. Lighter color indicates higher power/voltage. The blue and green lines indicate positions of the $\theta_H=0°$ spectra from Fig. 3a. The red curve is the '*cos*' model fit.

**Figure 6**. The angular dependence of the resonance field: Kittel's model (solid curve), '*cos*' model (dashed curve), and experimental data (red symbols).

**Figure 7**. Angular dependence of FMR in Fe foil. 2D gray-scale plot shows the absorption spectra as a function of $\theta_H$. Lighter color indicates higher power. The red curve is the '*cos*' model fit. The insert shows experimental schematic: a 0.1 mm thick Fe foil (1.3 mm × 0.7 mm) with an in-plane magnetic field applied at an angle $\theta_H$ with the rf current.

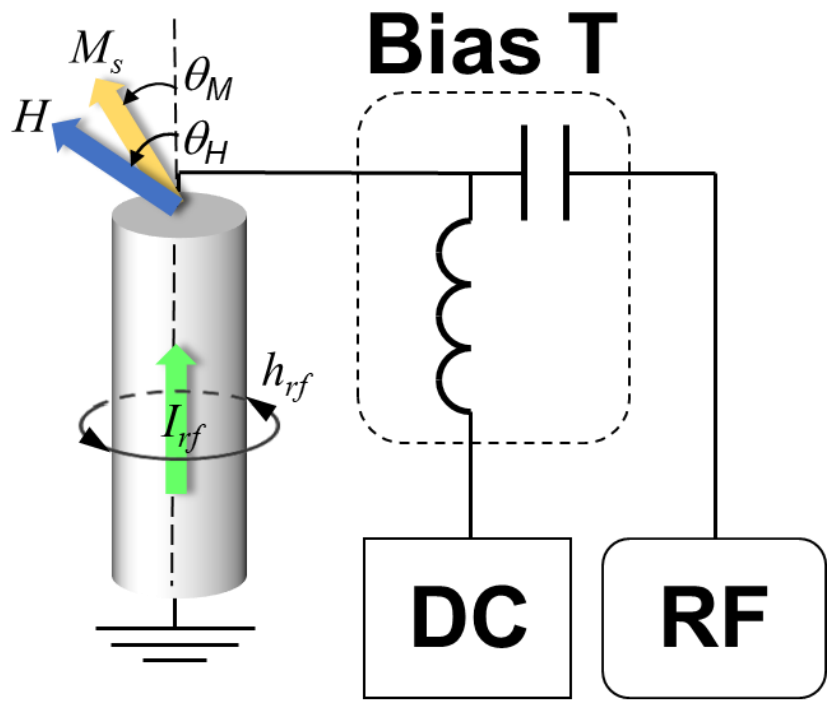

**Fig. 1**

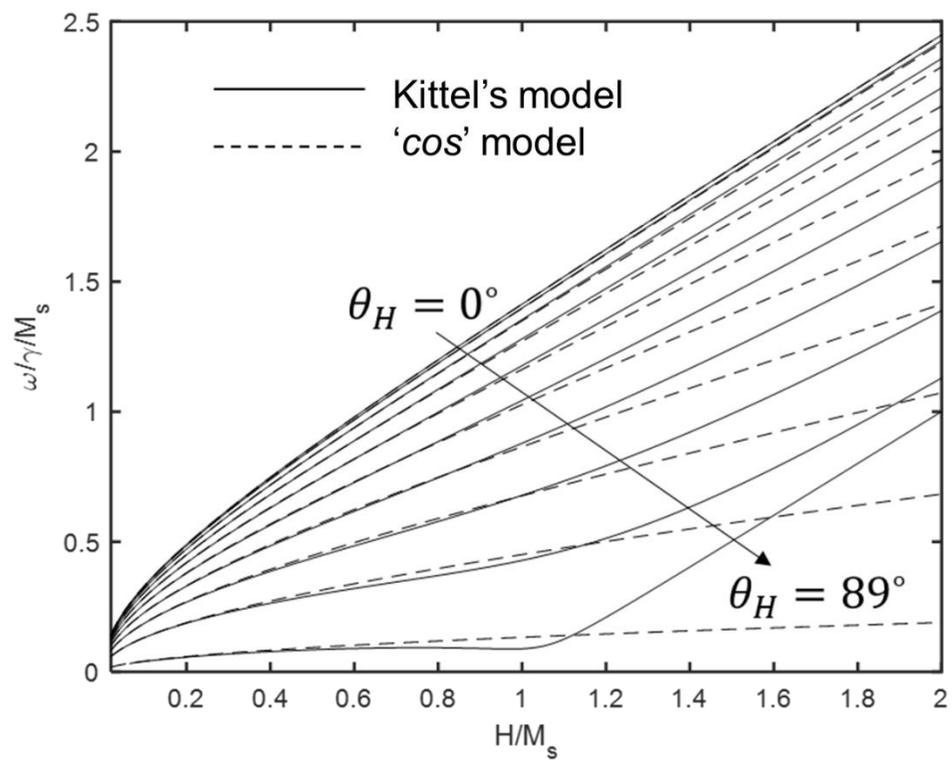

**Fig. 2**

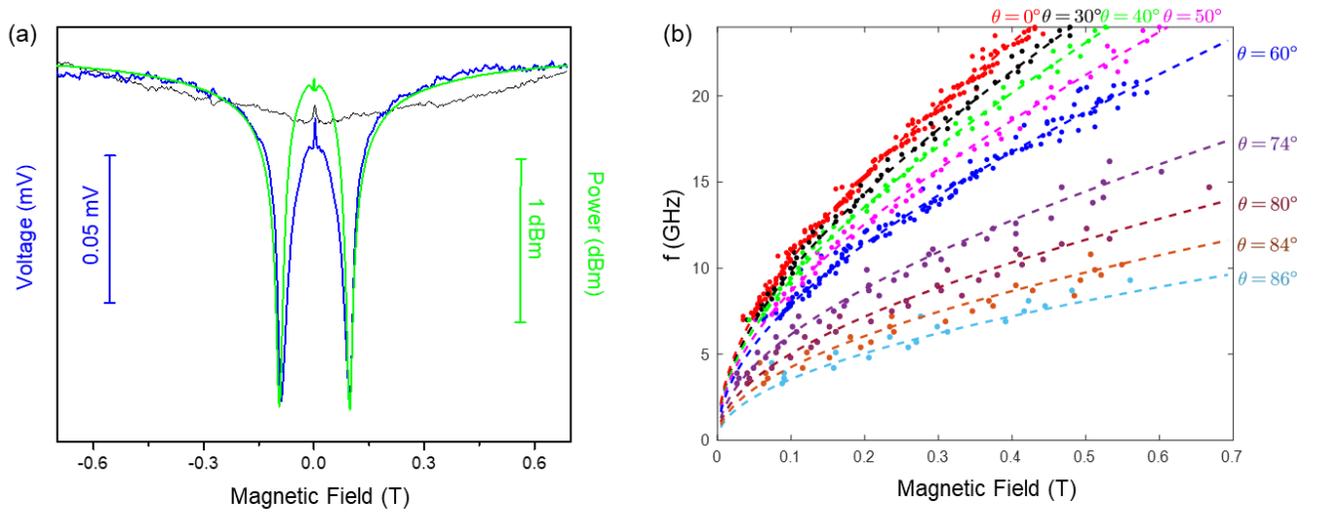

**Fig. 3**

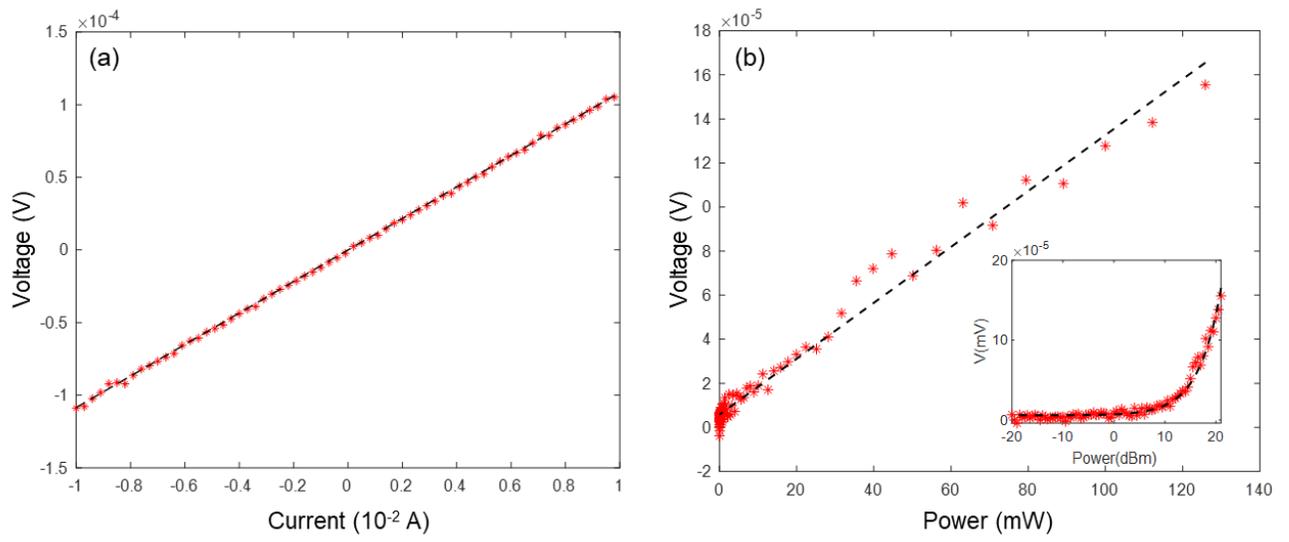

**Fig. 4**

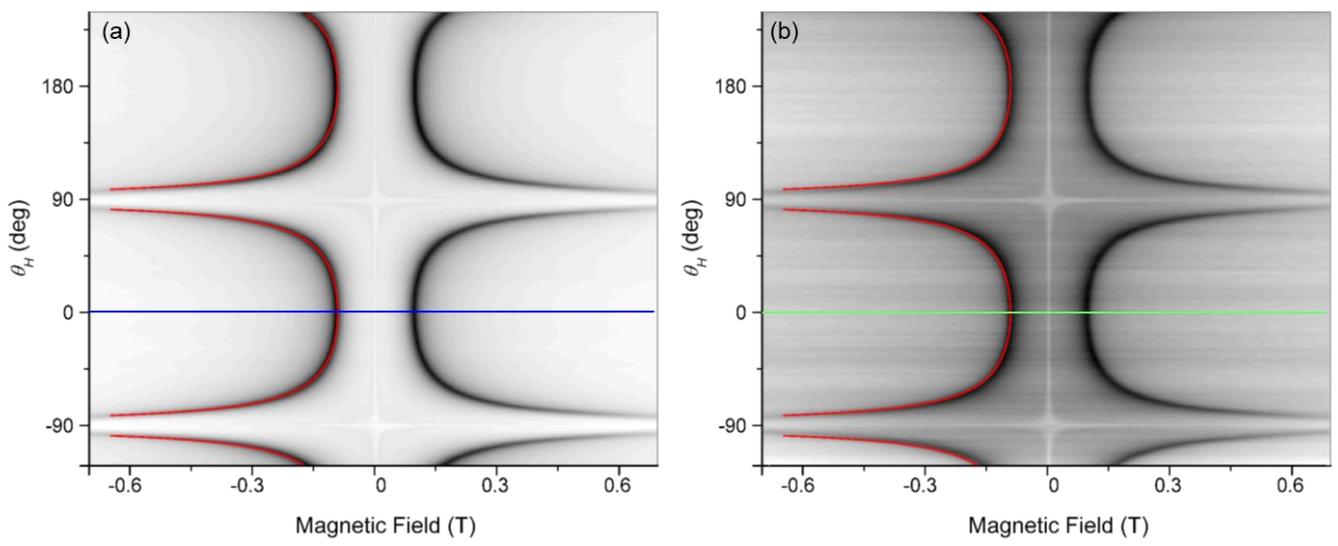

**Fig. 5**

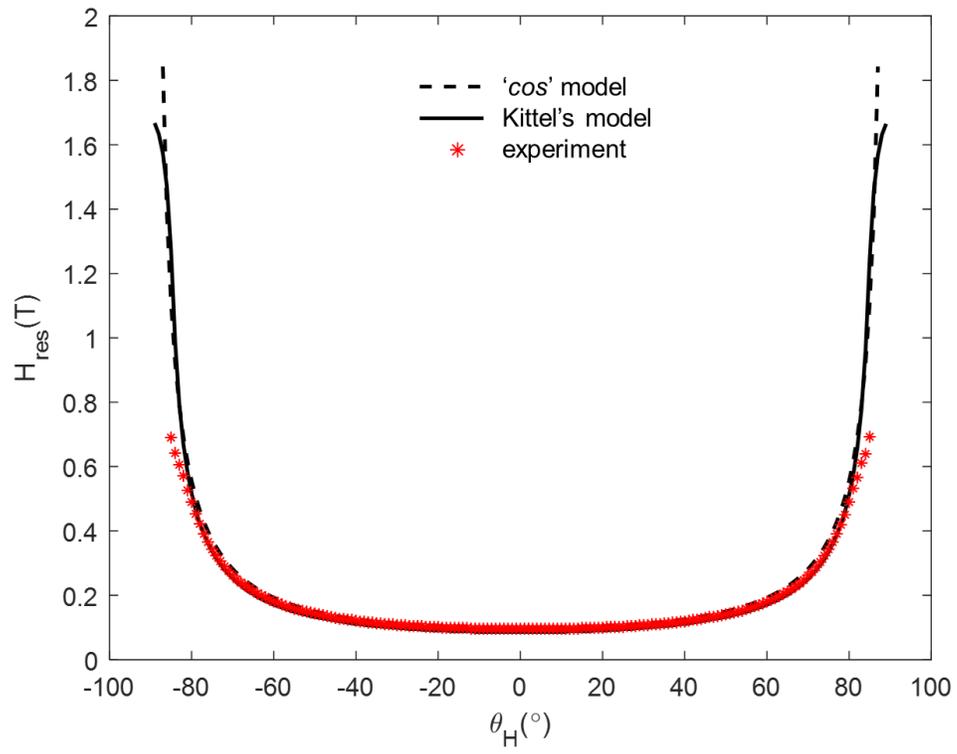

**Fig. 6**

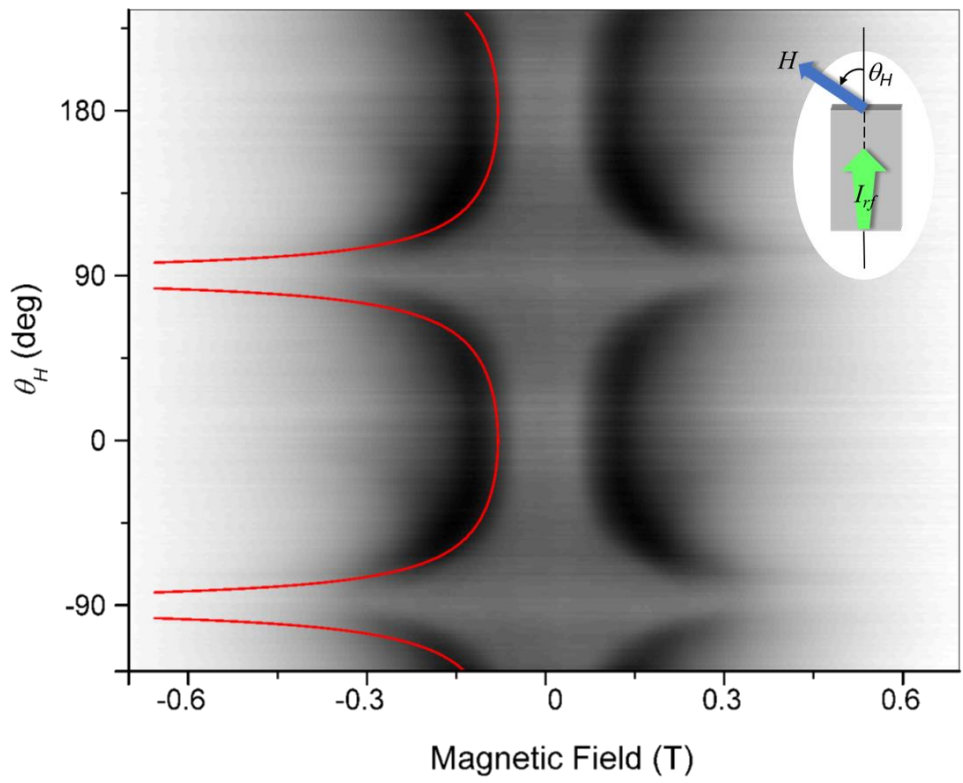

**Fig. 7**